\documentclass[english,aps,prb,preprint,superscriptaddress]{revtex4}
\usepackage[T1]{fontenc}
\usepackage[latin9]{inputenc}
\usepackage{amsmath}
\usepackage{graphicx}
\usepackage{amssymb}
\usepackage{esint}

\makeatletter

\newcommand{\lyxmathsym}[1]{\ifmmode\begingroup\def\b@ld{bold}
  \text{\ifx\math@version\b@ld\bfseries\fi#1}\endgroup\else#1\fi}

\@ifundefined{textcolor}{}
{%
 \definecolor{BLACK}{gray}{0}
 \definecolor{WHITE}{gray}{1}
 \definecolor{RED}{rgb}{1,0,0}
 \definecolor{GREEN}{rgb}{0,1,0}
 \definecolor{BLUE}{rgb}{0,0,1}
 \definecolor{CYAN}{cmyk}{1,0,0,0}
 \definecolor{MAGENTA}{cmyk}{0,1,0,0}
 \definecolor{YELLOW}{cmyk}{0,0,1,0}
 }

\usepackage{dcolumn}\usepackage{bm}\@ifundefined{definecolor}
 {\usepackage{color}}{}
\usepackage{subfigure}\@ifundefined{definecolor}
 {\usepackage{color}}{}
\usepackage{cancel}
\usepackage{times}

\makeatother

\usepackage{babel}

\begin{document}

\title{Bond Model and  Group Theory of Second Harmonic Generation in GaAs(001)}

\email[email: ]{hendradi_h@yahoo.com}

\author{Hendradi Hardhienata\footnote{on leave from Theoretical Physics Division, Bogor Agricultural University,  Jl. Meranti S, Darmaga, Indonesia}}
\affiliation{Center for surface- and nanoanalytics, Johannes Kepler University, Altenbergerstr. 69, 4040 Linz, Austria}
\author{Adalberto Alejo-Molina}
\affiliation{Centro de Investigac\'{i}on en Ingenier\'{i}a y Ciencias Aplicadas, UAEM Cuernavaca, Mor. 62160, Mexico}
\author{Andrii Prylepa}
\affiliation{Christian Doppler laboratory for microscopic and spectroscopic material characterization, Johannes Kepler University, Altenbergerstr. 69, 4040 Linz, Austria}
\affiliation{Center for surface- and nanoanalytics, Johannes Kepler University, Altenbergerstr. 69, 4040 Linz, Austria}
\author{Cornelia Reitb{\"o}ck}
\affiliation{Christian Doppler laboratory for microscopic and spectroscopic material characterization, Johannes Kepler University, Altenbergerstr. 69, 4040 Linz, Austria}
\affiliation{Center for surface- and nanoanalytics, Johannes Kepler University, Altenbergerstr. 69, 4040 Linz, Austria}
\author{David Stifter}
\affiliation{Christian Doppler laboratory for microscopic and spectroscopic material characterization, Johannes Kepler University, Altenbergerstr. 69, 4040 Linz, Austria}
\affiliation{Center for surface- and nanoanalytics, Johannes Kepler University, Altenbergerstr. 69, 4040 Linz, Austria}
\author{Kurt Hingerl}
\affiliation{Center for surface- and nanoanalytics, Johannes Kepler University, Altenbergerstr. 69, 4040 Linz, Austria}

\begin{abstract}
A comprehensive analysis of second harmonic generation (SHG) in a diatomic zincblende crystal based on the Simplified Bond Hyperpolarizability Model (SBHM), Group Theory (GT) is presented. The third rank tensor between the two approaches are compared and applied to reproduce rotational anisotropy (RA) SHG experimental result. It is well known that such a crystal is noncentrosymmetric, therefore the second harmonic generation source is dominated by bulk dipoles with a SHG polarization $P_{i}=\chi_{ijk}^{(2)}E_{j}(\omega)E_{k}(\omega)$.  We show that the SBHM previously applied to silicon can also be applied to an atomic cell containing two different atoms e.g. GaAs by introducing an effective hyperpolarizability $\alpha_{2GaAs}=\left(\alpha_{2Ga}-\alpha_{2As}\right)$ . Comparison with GT  yields the same third rank tensor for a $T_d$ point group corresponding to bulk tetrahedral structure.  Interestingly, SBHM gives good agreement with RA-SHG experiment of GaAs using only one single fitting parameter when Fresnel relations is also incorporated in the model.

\end{abstract}

\date{\today}

\maketitle

\section{Introduction}
The field of nonlinear optics flourished since the discovery of higher harmonic generation by Franken and coworkers in the 1960 [1].  Theoretical efforts to understand nonlinear phenomena soon followed, perhaps most notably the rigorous analysis of higher harmonics  by the group of Pershan [2]. The theory has been further developed by Bloembergen et al. [3] to explain nonlinearity in centrosymmetric material where it is argued that due to inversion symmetry nonlinear dipole contribution inside the material bulk is forbidden but electric quadrupole and magnetic dipole arising from quantum perturbative treatment of the interacting Hamiltonian are allowed. Other researchers apply parity consideration to  that the symmetric potential inside the bulk of a centrosymmetric material requires electronic transition from either symmetric-to-symmetric or antisymmetric-to-antisymmetric states which are forbidden thus resulting in zero components for the even order susceptibilty tensors and as a logical consequence even order nonlinear harmonics cannot exists inside the bulk [4]. 

However, it is interesting to note the work of Guidotti and Driscoll [5]. Based on their experiment, they argue that intense light illumination may produce dipolar contribution inside the bulk in contradiction with previous believes. Their argument is based on tensorial analysis and that although second harmonic generation intensity in Si(001)  experiment can be explained by fitting bulk quadrupole contribution the authors state that this agreement is maybe accidental. Other researchers rebutted this conclusion and, using symmetry considerations of the nanosecond lasers SHG experiment on Ge and Si semiconductor, argue that surface dipolar and bulk quadrupolar are contributing to the SHG intensities [6].

Since then, dipole contribution inside the bulk of a centrosymmetric crystal has been mainly neglected. Instead notable work particulary by Mizrahi and Sipe [7] was performed to develop the nonlinear polarization term in Ref. [3] using Fresnel equation. Their idea was soon developed further to analyze second harmonic (SHG) and third harmonic generation (THG) from vicinal Si(111) surfaces [8] yielding excellent fits but unfortunately requiring more than 10 input parameter that are perhaps not really independent of each other to reproduce experimental data. Unfortunately, the high amount of fit parameters makes the physical insight of nonlinearity in such systems more complicated.  Nevertheless, the general view holds that both the surface dipolar and bulk quadrupolar effects are both significant in generating correct SHG intensity profiles that are observed in reflection. Newer experimental techniques [9] suggest that magnetic dipolar effects to SHG can also not be neglected.

Although most of the work on SHG from centrosymmetric material is based on the phenomenological theory where the final formula is in the form of a Fourier series, another approach called the Simplified Bond Hyperpolarizability Model (SBHM) which is based on a classical Lorentz nonlinear oscillator model has been proposed in 2002 by Powell and coworkers [10]. Their effort was motivated by the complexity arising from phenomenological theory and notion that SHG experimental intensities from centrosymmetric system particularly vicinal Si(111) should be reproducable by another model using fewer parameters. Inspired by Ewald-Oseen's method [11,12] to calculate far field intensities in linear optics via superposition of dipoles, they extend this principle to nonlinear optics and obtain surprisingly a simplier analysis than the linear case [13].  The core idea of SBHM is that the driving field oscillates the charge in the material surface/interface harmonically and anharmonically along the bond direction between the atoms and because in classical electrodynamics an accelerated charge radiates, it produces linear and nonlinear harmonics in the reflection spectra. The advantage of using this simplified model is that -although it is not an ab initio theory- it gives a clear picture of the microscopic physics that produces higher harmonics. In addition, it requires less fitting parameters especially for low symmetry systems such as non vicinal centrosymmetric surfaces.

Since the model has been proposed, SBHM has achieved some  success. Assuming only contribution from the surface, SBHM was able to reproduce the four polarization SHG intensities of a Si(111) facet mentioned in Ref. [8] using only two parameters which are the complex up ($\alpha_u$) and down ($\alpha_d$) hyperpolarizabilities. However, it has also several limitations in that it could not fit certain Si orientation such as Si(011) and because it is generally seen as a surface model it cannot explain bulk related nonlinear source which has brought several criticism notably by McGilp [14]. As a response  Aspnes and coworkers extended their model to cope with retardation (RD), spatial-dispersion (SD), and magnetic (MG) effects and using this new contributions, they were able to fit the Si(011) data [15]. Other work was also performed by Kwon and coworkers [16] to explain SHG in non and vicinal Si(001) facet by introducing additional assumption namely that the anharmonic oscillation of the charge can also oscillate perpendicular to the bonds. They also argue that bulk quadrupolar contribution can be modelled in analogy to the surface dipolar fashion but using the multipole expansion of the nonlinear polarization or perturbative expansion form of the nonlinear polarization such as in [3] applying not three but four outer products of the bonds and one additional gradient vector representing the quadrupolar nature of the source. This bulk quadrupole  formulation is different than proposed by Peng [17] because the latter uses an expansion of the driving field gradient.       

In addition to SHG from centrosymmetric structures such as Si,  some work has been performed to investigate SHG in GaAs [18-22] which has a zincblende structure. Such a system is - in contrary to Si -  not centrosymmetric due to the different atomic potentials between Ga and As.  Therefore SHG produced by dipole inside the bulk is not forbidden and can be studied using the bond model which so far has been sucessful in predicting dipole radiation from centrosymmetric surfaces. Although the 4 fold symmetry has been obtained from various experiments, the exact origin is unknown [18]. Further investigation on vicinal GaAs structure also exist where the SHG contribution seems to arise from surface and bulk dipoles [19]. An experiment to measure  fourth harmonic generation from GaAs has been performed and found a strong 4 fold anisotropic surface specific signal that is not present in SHG  [20]. In addition SHG analysis from induced electric currents due to assymetric distribution of carriers in GaAs has been investigated using quantum mechanical arguments involving the creation of an exciton and band transition analysis [21]. 

To the best of our knowledge however, the classical bond model has never been applied  by other researchers to analyze SHG from a diatomic unit cell, in this case GaAs.  In our previous work [22] we have briefly discussed the relation between SBHM and group theory for diamond monocrystal and diatomic zincblende crystal. In this work we elaborate our points more clearly and go further by explaining how SBHM can really reproduce experimental results using only one fitting parameter and describe in length the mechanism and physical assumption that lead to the final SHG intensity formula and how it should be intepreted.

For clarity we organize this work as follows: we start by describing the instrumental setup of our experiment in Section II. We then proceed into Section III by introducing the force equation followed by  defining the corresponding bond direction for the unit cell and how it is inserted into the nonlinear SHG polarization. In Section IV we compare the model with the point group symmetry of diatomic zinc blende and check whether the two apporaches yields the same susceptibility tensor. Section V  gives a complete expression for the 4 polarization SHG intensity of GaAs that is obtained using SBHM. The experimental result is then fitted with the additional requirement of Fresnel equation as an important incorporation to the model. Finally, a brief summary describing the major points is presented.

\section{ Rotational Anisotropy SHG Experimental Setup}

Here we present our experimental setup and explain how the SHG intensity curve is obtained. Rotation anisotropy SHG (RA-SHG) is a technique that is able to monitor the crystal structure and the microgeometry of the surface of centrosymmetrical mediums which is often not available using linear diagnostic optical methods.

\begin{figure}[h]
\centering
\includegraphics[width=0.9\textwidth]{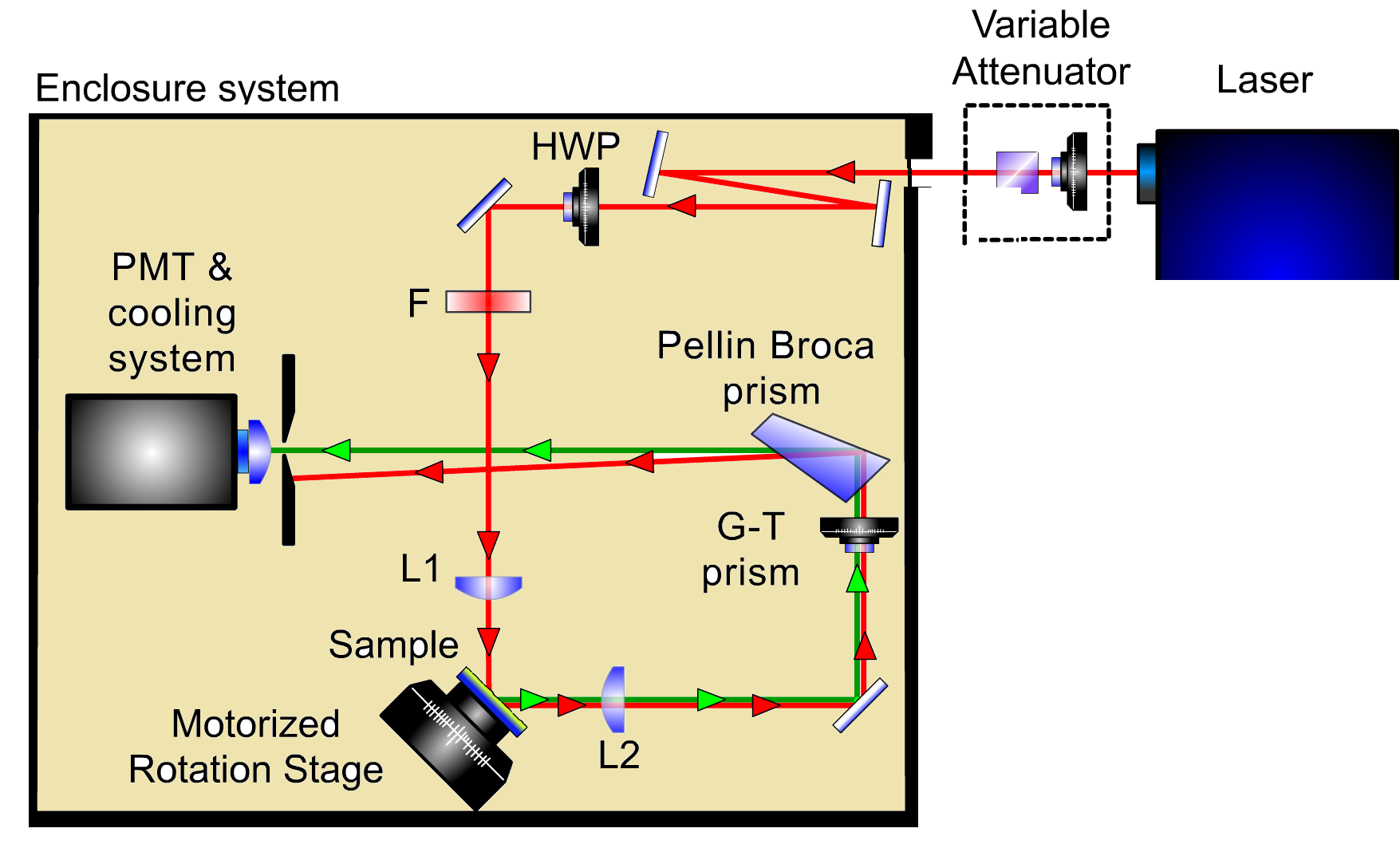}
\caption{Scheme of the RA- SHG setup. The following notations are used: HWP -- half-wave plate, PMT-- photomultiplier tube, G-T polarizer -- Glan-Taylor calcite polarizer, F -- filter, L1, L2 -- microscope objectives.}
\label{fig:setupPolarimeter}
\end{figure}

The schematic configuration of the experimental setup for RA-SHG measurements is shown in Fig. \ref{fig:setupPolarimeter}.
A diode pumped Nd:LSB solid-state sub-nanosecond (0,6 ns) pulse laser with wavelength of $1064$ nm was used as a source of radiation.
The fundamental radiation goes out of the laser and is passed through the power variable attenuator. After that the beam is directed to the half-wave plate which  determines the orientation of the plane of polarization of the incident beam.
Passing through filter (F) ensures that only laser light at fundamental wavelength is  focused on the sample by the lens. The beam is then focused onto the sample by the lens $ L1 $ with focal distance 25 mm that provides a spot diameter of $ \approx 10 \; \mu m $ in the focal position.

The incident angle of the fundamental radiation is $  45^ \circ  $ with respect to the surface normal.  
 The sample is placed on the motorized rotation stage which allows rotating the sample in the azimuthal plane.
The reflected SHG and the fundamental radiations are collected by the second lens $ L2 $ and are directed towards the Glan-Taylor calcite polarizer which is used as an analyzer of the polarization state of the SHG radiation. 
After that the radiation is directed to the Pellin-Broca prism near Brewster's angle minimizing reflection losses in $p$-polarized light. 
Then, the radiation exits at $  90^ \circ  $ with respect to the input direction plus a small angle due to dispersion which is less for longer wavelengths.

Due to this, SHG and residual fundamental radiations exit at different angles. 
The SHG radiation is passed to the photomultiplier tube (R7518, Hamamatsu) having a high sensitiveness at wavelength $ \lambda_{SHG}=532$  nm through the slit, which blocks the residual fundamental radiation. 
At the output of PMT, electric signal is gained by lock-in-amplifier SR830 (Stanford Research system).
The gained electric signal is recorded in the memory of computer by an acquisition card PCI-6110 from "National Instrument". 
The process of measurement is performed in automatic mode under control of personal computer and software "Labview". The average power of the incident radiation in the focal plane was set 35 mW that is lower than the damage threshold.

\section{ Bond Model for SHG in zincblende diatomic crystal}

SBHM is a classical model that in its most simplified form takes the following assumptions: The nonlinear radiations are only produced by anharmonic dipoles that are oscillating along the bonds.  Before proceeding further we would like to point out various sources that can generate SHG inside the material.  For the case of a centrosymmetric structure such as Si, dipole radiation inside the bulk is forbidden due to the fact that a symmetric potential leads to zero even order hyperpolarizability. However at the surface there is anisotropy and the symmetry is broken, therefore SHG can be generated from dipoles at the surface [23]. Also because there is a field gradient due to absorption SHG can also be generated inside the bulk by quadrupolar effects.

However, in a noncentrosymmetric crystal SHG dipole radiation is now allowed inside the bulk[24]. For GaAs this noncentrosymmetricity is due to the existence of two different atoms that sit in the center of each domain. It will be shown later that for this case the hyperpolarizability is not canceling out.  Furthermore because the driving field can of course experience significant decay there can be bulk quadrupolar effects too whose strength depend on the field decay gradient and penetration depth of the crystal. However if we assume that the field decay is sufficiently low there is not much difference between the fields strength at the layers below the first bulk layer and bulk dipoles then become the sole dominating SHG source. In this work we limit ourself to bulk dipoles only and show later on that this assumption can fit RA-SHG experiment nicely.

The GaAs atomic structure inside the bulk  is zincblende. Each atom is surrounded by 4 bonds forming a tetrahedral structure and has a $sp^3$ atomic orbital. The highest probability density of finding an electron is along the bond. Thus if the driving force has higher electric field component that is parallel in the direction of the bond it will produce a higher SHG contribution than if the field is aligned more perpendicular to the bond. Furthermore, the strength of the oscillation also depends on how easy the charges are moved by the driving field relative to the heavy nuclei which we assume at rest thus forming a static lattice. Therefore the electron nonlinear radiation also depends on the type of atom at the center of the tetrahedral cell. In the model, this effect is incorporated by specifying a different hyperpolarizability  for the Ga and As atom. 

\begin{figure}[hpbd]
\begin{center}
\includegraphics[scale=0.5]{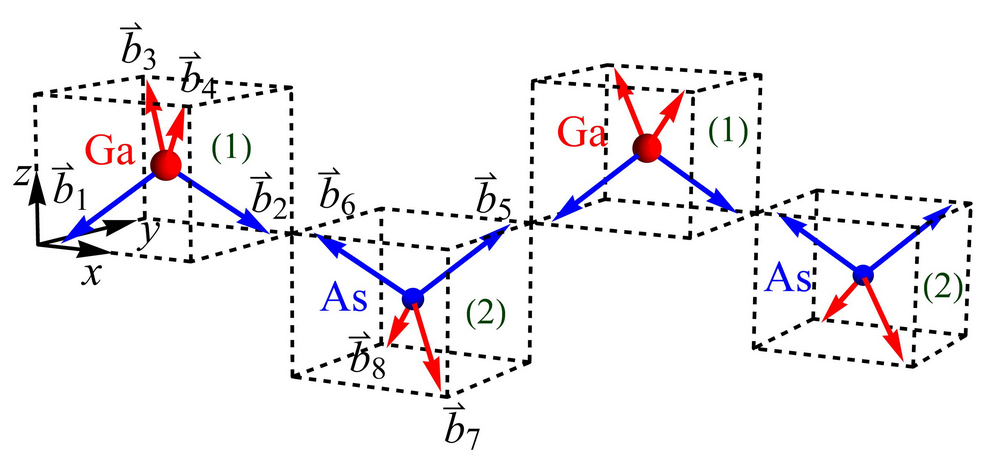}
\end{center}
  \caption{GaAs (001) configuration inside the bulk showing the bond direction relative to the coordinate system. The vector is defined by taking the center of the domain as the (0,0,0) coordinate. Each cube represents a domain consisting of four bonds. The system may freely rotate  around the $z$-axis.}
  \label{fig:potential}
\end{figure}

In the (001) direction the configuration of the bonds is depicted in Fig. 2 repeating itself along the $z$ direction. The choice of the coordinate system is arbitrary but we choose it in such a way so that it corresponds to the coordinate system choosen for comparison with group theory. To  analyze the symmetry more clearly we define two atomic domains labelled 1 and 2 each consisting of four bonds with two bonds directed upwards and two downwards but a different atom sitting in the center. Here we take for the first domain Ga as the center and for the the second domain As as the center.  If the two centers happened to be the same atom as in the case of Si we would have a  center of inversion that lies in the diagonal ends between the two domains. 

 The definition of the bond orientation in domain 1 is as follows 
\begin{equation}
\hat{b}_1=
\left(
\begin{array}{c}
-\frac{1}{\sqrt{2}}\sin\frac{\beta}{2}\\-\frac{1}{\sqrt{2}}\sin\frac{\beta}{2}\\-\cos\frac{\beta}{2}
\end{array}
\right)
\quad
\hat{b}_2=
\left(
\begin{array}{c}
\frac{1}{\sqrt{2}}\sin\frac{\beta}{2}\\\frac{1}{\sqrt{2}}\sin\frac{\beta}{2}\\-\cos\frac{\beta}{2}
\end{array}
\right)
\quad
\hat{b}_3=
\left(
\begin{array}{c}
-\frac{1}{\sqrt{2}}\sin\frac{\beta}{2}\\\frac{1}{\sqrt{2}}\sin\frac{\beta}{2}\\\cos\frac{\beta}{2}
\end{array}
\right)
\quad
\hat{b}_4=
\left(
\begin{array}{c}
\frac{1}{\sqrt{2}}\sin\frac{\beta}{2}\\-\frac{1}{\sqrt{2}}\sin\frac{\beta}{2}\\\cos\frac{\beta}{2}
\end{array}
\right)
\end{equation}
where $\beta=2\arccos[1/\sqrt{3}]\approx109.4^{0}$ is the angle between each bond. Meanwhile, a closer inspection to the geometry of the system quickly reveals that the bond orientation in the second domain is the same as in domain 1 but with a rotation of 90 degree about the z axis or about the [110] plane. Thus the bond orientation in domain 2 can be written as
\begin{equation}
\hat{b}_5=-\hat{b}_1\quad\hat{b}_6=-\hat{b}_2\quad\hat{b}_7=-\hat{b}_3\quad\hat{b}_8=-\hat{b}_4
\end{equation}

\begin{figure}[hpbd]
\begin{center}
\includegraphics[width=9cm]{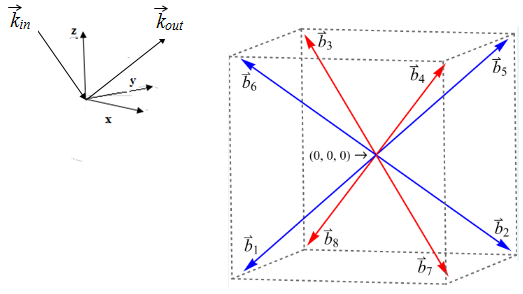}
\end{center}
  \caption{Bond vector orientation of the two domains centered at (0,0,0)}
  \label{fig:potential}
\end{figure}

Because each of the bond contribute SHG we must sum up over all eight bonds to obtain the total SHG radiation. It has to be noted that in the model all the bond vectors are defined relative to one center as depicted in Fig.3. 
This does not change the physics in the sense that nonlinear polarization wavelength is far larger than the domain distance so that the driving field that is experienced by each domain is practically the same. The hyperpolarizability of the first 4 bonds is the same but differs from the last four bonds, thus they are not cancelling each other out as in the case of Si(001) where the hyperpolarizability inside the bulk is the same for all 8 bonds.

Classically, nonlinearity can be seen as produced by a nonlinear polarization source term inside the material. This nonlinear source term is due to anharmonic oscillatory motion of the bound charge that is driven by the fundamental field. This anharmonic oscillation in turn produces a nonlinear radiation which can be detected in the far field. As noted by Aspnes in  Ref. [27]  this method is inspired by the Ewald-Oseen extinction theory where it has been shown in Ref. [28] that the formulas for reflection and transmission in linear optics can be derived microscopically within a classical framework in agreement with the macroscopic derivation using Maxwell boundary conditions. Therefore an extension into nonlinear optics by summing the nonlinear dipole contribution is a logical and ingenious step.  For the sake of clarity and comprehention, we derive the microscopic expression for the nonlinear polarization source starting from the nonlinear force equation taking the equilibrium position as zero [10, 29]:
\begin{equation}
F=q_{j}\vec{E}\hat{b_{j}}e^{-i\omega t}-\kappa_{1}x-\kappa_{2}x^{2}-\gamma\dot{x}=m\ddot{x}\end{equation}
here  $q, m, x$  are the electron charge, mass, and its displacement from equilibrium,  respectively and $\kappa_{1}$ and $\kappa_{2}$ are the harmonic and anharmonic spring constants, and the term $-\gamma\dot{x}$  is the common frictional loss in oscillation. Solving for $\triangle x_{1}$ and $\triangle x_{2}$ by using the assumption that $x$ can be written as $x=x_{0}+\triangle x_{1}e^{-i\omega t}+\triangle x_{2}e^{-i2\omega t}$ gives for the lowest order of approximation:
\begin{equation}
\triangle x_{1}=\frac{\hat{b_{j}}\cdot\vec{E}}{\kappa_{1}-m\omega^{2}-i\gamma\omega}
\end{equation}

\begin{equation}
\triangle x_{2}=\frac{\kappa_{2}}{\kappa_{1}-4m\omega^{2}-i\gamma2\omega}\left(\frac{\hat{b_{j}}\cdot\vec{E}}{\kappa_{1}-m\omega^{2}-i\gamma\omega}\right)^{2}\end{equation}
defining the linear and second order polarization that are produced by each bond $b_{j}$ as $p_{1j}=q_{j}\triangle x_{1}=\alpha_{1j}\left(\hat{b}_{j}\cdot\vec{E}\right)$ and $p_{2j}=q_{j}\triangle x_{2}=\alpha_{2j}\left(\hat{b}_{j}\cdot E\right)^{2}$
where $\alpha_{1}$ and $\alpha_{2}$ are the microscopic polarizability and second order hyperpolarizability. Because there are two atoms we have two linear susceptibilites $\alpha_{1Ga}$ and $\alpha_{1As}$ as well as two hyperpolarizabilities $\alpha_{2Ga}$ and $\alpha_{2As}$ .

In our experiment, the sample is rotated about the z axis so we need to define a  more general bond direction by allowing rotation along the [011] plane 
\begin{equation}
\hat{b}_{j}(\phi)=R_{z}(\phi
) \cdot \hat{b}_{j}\end{equation}

where  $R_z(\phi)$ is the common rotation matrix
\begin{equation}
R_z(\phi)=
\left(
\begin{array}{ccc}
\cos\phi&-\sin\phi&0 \\
\sin\phi&\cos\phi&0 \\
0&0&1
\end{array}
\right)
\end{equation}

The total linear and second harmonic polarization produced by all the bonds inside the bulk allowing rotation are thus :
\begin{equation}
\vec{p_{1}}=\frac{1}{V}{\sum\limits_{j=1}^{n}}\alpha_{1j}\hat{b}_{j}(\phi)\left(\hat{b}_{j}(\phi)\cdot\vec{E}\right)\end{equation}
and
\begin{equation}
\vec{p_{2}}=\frac{1}{V}{\sum\limits_{j=1}^{n}}\alpha_{2j}\hat{b}_{j}(\phi)\left(\hat{b}_{j}(\phi)\cdot\vec{E}\right)^{2}\end{equation}
where $n=8$ if we use two domains and $n=4$ if we use one domain.  Because each domain contains different atoms at its center we have two polarizabilities and hyperpolarizabilities. Therefore the total linear polarizability from  the two domains is
\begin{equation}
\vec{p_{1}}=\frac{1}{V}{\sum\limits_{j=1}^{4}}\alpha_{1Ga}\hat{b}_{j}(\phi)\left(\hat{b}_{j}(\phi)\cdot\vec{E}\right)+\frac{1}{V}{\sum\limits_{j=5}^{8}}\alpha_{1As}\hat{b}_{j}(\phi)\left(\hat{b}_{j}(\phi)\cdot\vec{E}\right)\end{equation}
whereas the total SHG polarizability is 
\begin{equation}
\vec{p_{2}}=\frac{1}{V}{\sum\limits_{j=1}^{4}}\alpha_{2Ga}\hat{b}_{j}(\phi)\left(\hat{b}_{j}(\phi)\cdot\vec{E}\right)^{2}+\frac{1}{V}{\sum\limits_{j=5}^{8}}\alpha_{2As}\hat{b}_{j}(\phi)\left(\hat{b}_{j}(\phi)\cdot\vec{E}\right)^{2}\end{equation}
here we assume that the susceptibility and hyperpolarizability does not depend on the orientation of the $j-$th bond relative to the field but only depends on the type of the atomic center because they are inside the bulk

We can write the total bulk polarization in a more compact form
\begin{equation}
\label{eqn:P}
\vec{P}=\chi_{1Ga}\vec{E}+\chi_{1As}\vec{E}+\chi_{2Ga}\vec{E}\vec{E}+\chi_{2As}\vec{E}\vec{E}
\end{equation}
where $V$ is the volume and $\chi_1$ and $\chi_2$ are the first- and second-order suceptibility tensors of the system. In the preceding equation the nomenclature highlights already the tryadic product of the bond directions, as well the dyadic product of the electric fields. Although there are two different hyperpolarizabilities for the SHG polarization the experimenter cannot distinguish the SHG contribution from the first and second domain and thus can only measure the effective hyperpolarizability/susceptibility. In other words because each bond in a domain has its corresponding antibond with a negative bond vector in the other domain the total SHG radiation is the difference between the SHG radiation from the first and second domain. In other words due to the symmetric fact we can reduce the problem from eight bonds (2 domains) into four bonds (effective domain) by introducing the effective hyperpolarizability
\begin{equation}
\alpha_{2GaAs}= \left(\alpha_{2Ga}-\alpha_{2As}\right)\end{equation}
Therefore Eq. (13) for the SHG part now can be written as
\begin{equation}
\label{eqn:P}
\vec{P}_{SHG}=\frac{1}{V}{\sum\limits_{j=1}^{4}}\alpha_{GaAs}\hat{b}_{j}(\phi)\left(\hat{b}_{j}(\phi)\cdot\vec{E}\right)^{2}=\chi_{GaAs}^{(2)}\vec{E}\vec{E}
\end{equation}
where we have for our coordinate system the definition of the incoming unit vector for the fields:
\begin{equation}
\hat{E}_{p}=\left(\begin{array}{c}
\cos\theta_{i}\\
0\\
\sin\theta_{i}\end{array}\right)\end{equation}

\begin{equation}
\hat{E}_{s}=\left(\begin{array}{c}
0\\
1\\
0\end{array}\right)\end{equation}
The nonlinear third rank tensor $\chi_{GaAs}^{(2)}$ in Eq.(14) is obtained by direct product over all the bond directions. One might ask the question whether such an operation is a valid way in obtaining the tensor. Therefore, a comparison of the tensor from group theory is presented. 

\section{Group Theory analysis of Gallium Arsenide}
Group theory is a mathematical tool that can be applied to investigate the group of symmetry that belongs to a crystal. It is related to the physics by the so called  Neumann's Principle [22]: 

\begin{center} {}''The symmetry elements of any physical crystal property must include the symmetry elements of the crystal point group.{}'' \end{center}
In other words the optical properties such as the nonlinear susceptibility that are attributed to a certain crystal must contain the tensorial elements obtained from group theory. The latter is obtained by investigating all the allowed symmetric operations such as rotation and reflection for specific mirror planes. The allowed operations are those that do not change the physics or crystal properties and form a point group. 

Fortunately the symmetry groups  of the second and third rank tensor (or even fourth rank tensor) for several crystal orientations are well known and can be found in standard group theory literature such as Nye [25] or books that focus on group theory applications [26]. As can be seen from Fig. 4 GaAs conventional cell is a zincblende which can be generated using tetrahedral structures. We have shown in the previous section that inside the bulk the 8 bonds can be decomposed into a 4 bond tetrahedral with an effective susceptibility. This structure has a symmetry group called $T_{d}$ and has a third rank tensor in the form
\begin{equation}
\chi_{ijk}^{(2)}=\left(\begin{array}{c}
\left(\begin{array}{ccc}
0 & 0 & 0\\
0 & 0 & d_{132}\\
0 & d_{132} & 0\end{array}\right)\\
\left(\begin{array}{ccc}
0 & 0 & d_{132}\\
0 & 0 & 0\\
d_{132} & 0 & 0\end{array}\right)\\
\left(\begin{array}{ccc}
0 & d_{132} & 0\\
d_{132} & 0 & 0\\
0 & 0 & 0\end{array}\right)\end{array}\right)\end{equation}
where $d_{132}=\chi_{123}$ is an arbitrary constant that in this case must be fitted. It is interesting that already here GT demand only one arbitrary parameter for the fit. The effective third rank tensor describing SHG can be obtained from Eq. (14)  
\begin{equation}
\chi_{ijk}^{(2)}=\frac{\alpha_{2GaAs}}{V}\underset{j=1}{\overset{4}{\sum}}b_{j}(\phi)\otimes b_{j}(\phi)\otimes b_{j}(\phi)\end{equation}
The next step is to apply the same initial coordinate that is applied in the standard textbooks and those from the SBHM because in general the tensor elements contain a rotation matrix whereas in Ref. [25,26] they are fixed on the coordinate given in Fig. [4]. 

\begin{figure}[hpbd]
\begin{center}
\includegraphics{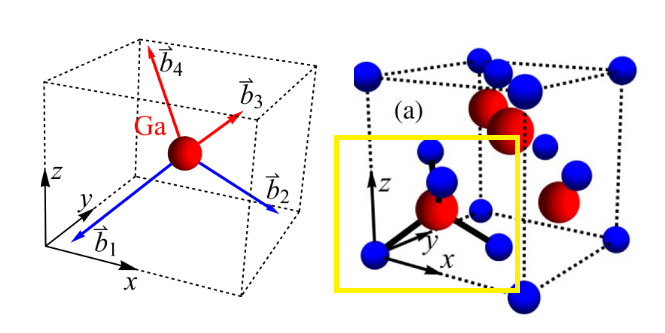}
\end{center}
  \caption{ (a) Definition of the SBHM GaAs (001) bond coordinate and (b) the configuration used in group theory textbook , Refs.[25,26].}
  \label{fig:potential}
\end{figure}

As can been seen it turns out that they are the same for this case because we have choosen the bond direction definition before in such a way so that it does not need to be rotated. This will not be the case for other orientations e.g. GaAs(111) where one has to perform two rotations for the comparison.  Therefore we can simply set $\phi=0$ in Eq. (18) and generate the third rank tensor which takes the form [22]:
\begin{equation}
\chi_{ijk}^{(2)}=\left(\begin{array}{c}
\left(\begin{array}{ccc}
0 & 0 & 0\\
0 & 0 & -\alpha_{GaAs}\sin\frac{\beta}{2}\sin\beta\\
0 & -\alpha_{GaAs}\sin\frac{\beta}{2}\sin\beta & 0\end{array}\right)\\
\left(\begin{array}{ccc}
0 & 0 & -\alpha_{GaAs}\sin\frac{\beta}{2}\sin\beta\\
0 & 0 & 0\\
-\alpha_{GaAs}\sin\frac{\beta}{2}\sin\beta & 0 & 0\end{array}\right)\\
\left(\begin{array}{ccc}
0 & -\alpha_{GaAs}\sin\frac{\beta}{2}\sin\beta & 0\\
-\alpha_{GaAs}\sin\frac{\beta}{2}\sin\beta & 0 & 0\\
0 & 0 & 0\end{array}\right)\end{array}\right)\end{equation}
which is exactly the same as the tensor in Eq.(17) obtained by group theory. Thus we have demonstrated that for the GaAs bulk SBHM and GT gives the same tensor elements and the same independent parameter which is the effective susceptibility. If there would have been a  different result one could have questioned the validity of SBHM in modelling the experiment because of violation of the Neumann's principle.

\section{Experimental Results and Discussion}

In this section we compare the experimental result from RA-SHG spectroscopy on GaAs with SBHM. With the assumption that the bond charges radiate as dipoles, the far-field radiation $E_{ff}$ can be written as [10,16,29]
\begin{equation}
\label{eqn:Eff}
\vec{E}_{ff}=k^2 \frac{e^{ikr}}{r}\left(\vec{P}-\hat{k}\left[\hat{k}\cdot\vec{P}\right]\right)=
k^2\frac{e^{ikr}}{r}\left(I-\hat{k} \hat{k}\right)\cdot\vec{P}
\end{equation}
where $k$ is the direction of the outgoing (observer) SHG wave and can be written for the $p$-input case
\begin{equation}
\hat{k}=\sin\theta_{o}\hat{x}+\cos\theta_{o}\hat{z}
\end{equation}

Eq. (20) can be evaluated to obtain the four polarization intensities $p$-in $p$-out, $p$-in $s$-out, $s$-in $p$-out, and $s$-in $s$-out:
\begin{eqnarray}
\mathbf{I}_{pp}&=\frac{4}{27}\alpha\alpha^{*}\sin^{2}2\phi\cos^{2}\theta_{i}\left(3\sin\left[\theta_{i}-\theta_{0}\right]+\sin\left[\theta_{i}+\theta_{0}\right]\right)^{2} &\end{eqnarray}
\begin{eqnarray}
\mathbf{I}_{ps}&=\frac{16}{27}\alpha\alpha^{*}\cos^{2}2\phi\sin^{2}2\theta_{i} \end{eqnarray}
\begin{eqnarray}
\mathbf{I}_{sp}&=\frac{16}{27}\alpha\alpha^{*}\sin^{2}2\phi\sin^{2}\theta_{o} \end{eqnarray}
\begin{eqnarray}
\mathbf{I}_{ss}&=0 \end{eqnarray}
where we have abbreviated $\alpha_{2GaAs}$ as $\alpha$.
From the SHG intensity formulas above it is straightforward to see that the produced azimuthal rotational SHG intensities (except for $ss$ which is zero) should be  4 fold due to the $\sin^{2}(2\phi)$ or $\cos^{2}(2\phi)$ term which also suggest a phase shift of  $\phi=45^o$ between the $ps$ and the other two polarizations, the latter two being in phase to each other. Physically for the given configuration the dipoles oscillating along the bonds radiate a stronger field when measured in the $s$-out direction rather than the $p$-out.  The most interesting feature however is that the intensity requires only one experimental parameter (see Eq. (19)) which is the effective hyperpolarizability or $\alpha_{2GaAs}$. 

\begin{figure}
\begin{center}
\includegraphics[width=14cm]{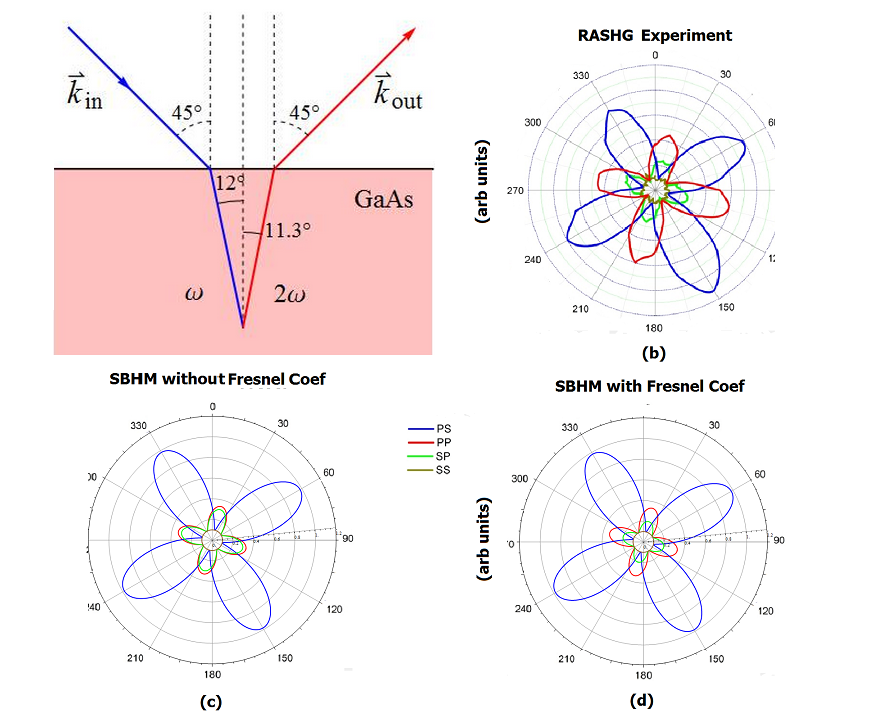}
\end{center}
  \caption{GaAs (001) experiment vs theory. (a) Sketch of the change of the driving field and SHG field direction inside the material. (b) RA-SHG experiment (c) SBHM simulation without Fresnel coefficients (d) SBHM simulation with Fresnel coefficient included.}
  \label{fig:potential}
\end{figure}

 Fig. (5) depicts the simulation for the incoming fundamental and outgoing SHG. In the experiment the incoming field is incident at $\theta_{i}=\theta_{o}=45^{0}$ but because it is the bulk dipole that radiates SHG driven by the fundamental field inside the bulk, the angle is adjusted using Snell's law by taking $n_{\omega GaAs}=3.4$ so that $\theta_{i}=12^{0}$ and because the outgoing wave is radiating at $2\omega$ therefore we use $n_{2\omega GaAs}=3.6$ resulting in $\theta_{o}=11.3^{0}$ (see the sketch in Fig 5a). The fitting result is seen in Fig. 5c and gives a good match for $\alpha_{2GaAs}=2.8$  where we put a dc factor $0.1$ to model the experimental noise (viewed as a circle around the center). 

What is not satisfied is that the $pp$ and $sp$ SHG radiation are likely in the same order which is contradictory to the experiment given in Fig 5b. However, we found that this problem can be resolved if we include the Fresnel coefficients which is incorporated in standard SHG phenomenological theory [7]  but as far as we know has never been implemented in SBHM. Using the well known Fresnel formulas with $n_{1}=1$ (air), $n_{2}=3.4$ (GaAs refractive index for $\omega$), and $n_{22}=3.6$ (GaAs for $2\omega$) we obtain for the $p$ and $s$ transmission fundamental coefficients from the air to the material:
\begin{equation}
t_{p}=\frac{2n_{1}\cos\theta_{o}}{n_{2}\cos\theta_{i}+n_{1}\cos\theta_{o}}=\frac{2(1)\cos12^{\circ}}{3.4\cos45^{\circ}+1\cos12^{0}}\thickapprox 0.58\end{equation}
\begin{equation}
t_{s}=\frac{2n_{1}\cos\theta_{i}}{n_{1}\cos\theta_{i}+n_{2}\cos\theta_{t}}=\frac{2(1)\cos45^{\circ}}{1\cos45^{\circ}+3.4\cos12^{0}}\thickapprox 0.35\end{equation}
whereas for the $p$ and $s$ SHG transmission from the material to air coefficents:
\begin{equation}
t_{2p}=\frac{2n_{22}\cos\theta_{o}}{n_{1}\cos\theta_{i}+n_{22}\cos\theta_{o}}=\frac{2(3.6)\cos11.3^{\circ}}{1\cos45^{\circ}+3.6\cos11.3^{\circ}}\thickapprox 0.43\end{equation}
\begin{equation}
t_{2s}=\frac{2n_{22}\cos\theta_{i}}{n_{22}\cos\theta_{i}+n_{1}\cos\theta_{t}}=\frac{2(3.6)\cos45^{\circ}}{3.6\cos45^{\circ}+1\cos11.3^{\circ}}\thickapprox 0.40\end{equation}
Because the nonlinear polarization is proportional to the square of the driving field the transmission coefficient of the fundamental from air to media has to be squared whereas the SHG radiation is proportional to the $2\omega$ field. Therefore for the $pp$, $ps$, and $sp$ one has the following Fresnel coefficients
\begin{equation}
f_{pp}=t_{p}t_{p}t_{2p}\thickapprox0.143\end{equation}
\begin{equation}
f_{ps}=t_{p}t_{p}t_{2s}\thickapprox0.136\end{equation}
\begin{equation}
f_{sp}=t_{s}t_{s}t_{2p}\thickapprox0.053\end{equation}
Therefore the $sp$ SHG intensity is about $\frac{f_{sp}}{f_{pp}}\thickapprox 0.37$ times smaller than the $pp$ polarization and this has to be accounted in the model. Also the $pp$ SHG intensity is just slightly larger than the $ps$ by a factor of $\frac{f_{pp}}{f_{ps}}\thickapprox 1.05$. Doing this adjustment one arrives at Fig. 5d this time with $\alpha_{2GaAs}=3$ which gives a good fit when compared to the RA-SHG data. However the $pp$ SHG polarization intensity is smaller than in the experiment. We attribute this due to neglect of possible surface interface contribution where there is perhaps an additional field along the $z$ direction which is clearly the case in Ref. [21] that contributing to the increase in $p$ out polarization, but this is still an open question and requires further investigation.

\section{Summary}
We have provided for the first time a comprehensive analysis of SHG in reflection from a diatomic zincblende structure e.g. GaAs involving the bond model, group theory, and state the importance of including Fresnel analysis in SBHM. We show by considering bulk dipole as the main SHG contribution in GaAs that the third rank hyperpolarizability tensor obtained by assuming anharmonic SHG dipole oscillation along the bond (SBHM)  is exactly equal with the tensor obtained from group theory (GT) which is the $T_d$ point group. Both SBHM and GT demand only one independent parameter in the form of an effective hyperpolarizability to fit the RA-SHG experiment. The model correctly predicts the phase and the 4 fold symmetry and gives a good agreement of the SHG intensity profile if Fresnel coefficients are further incorporated in SBHM. 

\emph{Acknowledgements}:  The authors would like to thank financial support by the Austrian
Federal Ministry of Economy, the Austrian Family and Youth and the Austrian National
Foundation for Research, Technology and Development. H.H. would also like to acknowledge funding from the Technology Grant Southeast Asia.

\nopagebreak

\bibliography{SHGSiReferencesEwald2}

\end{document}